# Remarks on entanglement and subjectivity..

Vittorio Cantoni
vittorio.cantoni@libero.it

ABSTRACT

It is maintained that the phenomenon of entanglement, as presented by Bohm, admits, for every observation, a *subjective interpretation* which is free from action at a distance or superluminal transmission of information.

**Introduction.**

In Einstein's words, "Physics should represent a reality in time and space, free from spooky action at a distance."[1] We show that, contrary to what is frequently asserted, the Quantum Entanglement admits for each observer a *subjective interpretation* which does not imply action at a distance or transmission of information of any kind. Therefore it is compatible with the structure of relativistic space-time and in agreement with Einstein's aforementioned belief. Though the subjective interpretations associated with distinct observations involve distinct (but unobservable) elements, they all lead to the same (complete and directly observable) objective description of the entanglement phenomenon

We shall confine the discussion to David Bohm's simplified version of the Einstein Podolsky Rosen (EPR) thought experiment[2], and shall only use well-known notions and well-established elementary facts about the quantum-mechanical description of spin ½ particles. In this simplified approach, the only observables that need to be considered are the spin observables, regarded, as usual in Quantum Mechanics[3], as completely characterizing the quantum states through the statistical distributions of the results of their repeated measurements on *ensembles* of copies of the states.

**1. One spin ½ particle in a pure state.**

First, consider a spin ½ particle *prepared*, by means of a suitable *source* (involving, for example, a Stern-Gerlach apparatus) in a state $s^*+$ with spin parallel to a given oriented direction $d^*+$, so that one has

$$p(s^*+,d^*+) = 1 \quad \text{and} \quad p(s^*+, d+) = cos^2[½ (d^*+,d+)],$$

where $p(s+,d+)$ denotes the probability that, for the generic state $s+$, the measurement of the spin component in the generic oriented direction $d+$ be positive, and $(d^*+,d+)$ denotes the angle between $d^*+$ and $d+$. (In our notation, substitution of the index "+" by "-" reverses the orientations of the spins and the directions, and $d$ denotes the common non-oriented direction of $d+$ and $d-$).

The measurement of the component of the spin of our particle in the state $s^*+$ along any oriented direction $d+$ can be performed in principle, by means of an *analyzer* (for example a second suitably oriented Stern-.Gerlach apparatus*)*. Only in the exceptional case in which the spin component is measured along the direction $d^*$ itself such a measurement leaves the particle in the original state $s^*+$ (a circumstance expressed by saying that the particle is not *perturbed* by the measurement). In all other cases the sign of the measured spin component is unpredictable, and has probability $p$ strictly positive and less than 1: in this case the measurement amounts to the preparation of a new state with spin parallel to $d+$ or to $d-$, according to the outcome. Thus, in general, the measurement performed on a single particle prepared in the state $s^*+$ gives practically no information about $s*+$ itself, i.e. about the state of the particle *before* the measurement. In order to acquire significant information about $s*+$ one would have to repeat the same measurement many times on the elements of an *ensemble* of particles all prepared in the state $s*+$, and derive from the results of the repeated measurements an estimate of the probability $p(s*+,d+)$. Except in the special cases in which $d$ is chosen coincident with $d*$, after a specific element of the *ensemble* has been used to perform the measurement of the spin in some direction $d$ the properties of the element are modified, so that no further information on the state $s*+$ can be gathered by subsequent measurements on what has become of that element.

---

[1] Walter Isaacson, *Einstein, his Life and Universe,* (Simon & Schuster 2008) ch. 20.
[2] A. Einstein, B. Podolsky and N. Rosen, "Can the Quantum-Mechanical description of Physycal Reality be considered Complete?", *Phys. Rev.* **47**, 777 (1935).
  David Bohm, *Quantum Theory* (Prentice hall Inc.1951 and Dover Publications Inc. 1989) ch. 22.18.
[3] See for example G. W. Mackey, *Mathematical Foundations of Quantum Mechanics,* (W.A. Benjamin Inc. 1963).

## 2. One spin ½ particle in the spherically symmetric state.

Next, consider the *spherically symmetric* state $s°$ of a particle of the same kind as the one just considered. By definition, $s°$ is characterized by the probability function

$$p(s°, d+) = ½ \text{ for every oriented direction } d+.$$

In contrast with the states of the kind of $s*+$ in the previous section, which was a *pure state*, $s°$ is a *mixture*[4]. In particular, it can be regarded as the mixture, with weights (½, ½), of two pure states $s*+$ and $s*-)$ with opposite spin directions $d*+$ and $d*-$, a fact expressed by the relation

$$p(s°, d+) = ½\, p(s(d*+), d+) + ½\, p(s(d*-), d+)$$

(because $p(s(d*+), d+) + p(s(d*-), d+) = cos^2[½\,(d*+,d+)] + cos^2[½\,(d*-,d+)] = cos^2[½(d*+,d+)] + sin^2[½\,(d*+,d+)] = 1$). Since the relation holds for every choice of the pair of states $(s*+, s*-)$, there are as many such decompositions as there are directions in space.

From the point of view of an observer $O*$ measuring the component in the direction $d*+$ of the spin of the elements of an *ensemble* of particles all prepared in the spherically symmetric state $s°$, the decomposition expressed by the last relation can consistently be interpreted as follows: the elements of the ensemble are regarded as pairs of particles in the pure states $s*+$ and $s*-$ with opposite spins in the direction $d*$, and the measurement picks one of them at random (with probability ½ for each outcome) and measures its spin component along $d*+$, thus identifying which of the two particles happened to be selected. Consistently with this interpretation the measurement of the spin component along $d*+$ does not perturb the pure state of whichever of the component particles $s*+$ or $s*-$ has been selected, while the measurement of the spin component in any other direction would alter it in an unpredictable way. The interpretation is essentially *subjective:* everything we have said could be repeated for any observer $O*'$ associated with a new direction $d*'$ distinct from $d*$, after replacement of the pair $(s*+, s*-)$ by the new pair of pure states $(s*'+, s*'-)$ with opposite spins in the new direction, so that *in no objective way can the spin ½ particle in the spherically symmetric state be regarded as a particle in some pure state of unknown spin direction*. However, the various subjective interpretations (each of which involves the choice of a specific but unobservable decomposition) all lead to the common (complete and observable by means of series of repeated experiments) *objective* characterization of the spherically symmetric state in terms of its defining probability function given above.

Also, from the repetition (many times) on particles in the state $s°$, of the two-step experiment consisting (step 1) in the observation of the sign of the component of the spin along a given direction $d+$, followed (step 2) by the observation of the sign of the spin of what has become of *that same particle* along a second given direction $d'+$, one can deduce an estimate of the probability $p(d,d')$ that the signs found at the two steps be both positive. Such a probability function $p(d,d')$ is *objective* by definition, and from the defining properties of the spherically symmetric state it is immediately seen to depend on the angle $(d+,d'+)$ according to the law

---

[4] See for example Mackey, op. cit.

$$p(d,d') = \tfrac{1}{2} \cos^2[\tfrac{1}{2}(d+,d'+)].$$

The same result turns out to be derivable by regarding the state $s°$ as a mixture of the two states $s*-$ and $s*+$, *i.e.* by adopting the subjective interpretation of the observer $O*$: however the role of the pair ($a*-, a*+$) in this derivation turns out to be purely auxiliary and does not appear in the final result. So everything goes "as if" the state $s°$ had been prepared as a mixture of the pair of pure states *$a*-$ and $a*+$* (or of any other pair of pure states with opposite spins), though in fact the actual preparation of $s°$ is undetermined and irrelevant.

### 3. Two spin ½ particles in entangled state.

Finally, let us consider the system of two spin ½ entangled particles occurring in the Bohm's version of the EPR thought experiment.

One has a source $S$ constituted of an *ensemble* of identical particles. Each particle of the ensemble is a spin zero particle, and spontaneously decays into two spin ½ mutually identical particles. By the laws of conservation of momentum and spin, at each decay the two particles produced move apart from $S$ with opposite directions, and constitute a two-particle system of total spin zero.

Two analyzers $A$ and $A'$ are placed at opposite sides with respect to the source $S$, so that $A$, $S$ and $A'$ are aligned. Whenever a pair of particles resulting from the same decay are emitted in a direction sufficiently close to the line $ASA'$, one of the particles, say $a$, is detected at $A$ and the other, $a'$, is detected at $A'$, after times of flight proportional to the distances of $A$ and $A'$, respectively, from the source.

It is convenient to distinguish three distinct systems

The first, that we shall call the *whole system*, is the two-particle system of the pair $(a,a')$ of decay products, which presents itself in a state $E$ that we shall call the *entangled state*, characterized by the preparation consisting in just waiting for a decay to occur in the direction leading to the detectors. The ensemble on which repeated measurements on the state $E$ can be performed is a set of pairs such as $(a,a')$, from successive decays.

The second system is the one-particle system of particle $a$ alone (or *partial system at A*, as we shall call it). It presents itself in a state whose preparation consists in waiting for a decay to occur, and observing what happens at $A$ (ignoring the very existence of particle $a'$). The defining ensemble of this state is therefore a set of successive particles arising from decays at $S$ and detected at $A$. Since, as far as the spin components are concerned, there is no preferred direction in the decay process, the experiment produces the particle in the spherically symmetric state $s°$ of the previous section, because in every oriented direction $d+$ the analyzer $A$ has the same probability ½ of measuring a positive spin component.

The characterization of the third system (or *partial system at A'*) is obtained from the preceding paragraph simply by interchanging $a$ with $a'$, and $A$ with $A'$. Consequently the state $s°'$ of the particle $a'$ prepared by the experiment is also spherically symmetric.

By the very description of their preparations, the whole system and the two partial systems are related by the following properties:

1) To each detection of a particle $a$ at $A$ corresponds the possible detection of a particle $a'$ at $A'$, so that in repeated experiments the pairs of corresponding detections can be labeled and recognized.
2) The conservation of the total spin imposes that if the detectors at $A$ and $A'$ are set to measure the spin components of the incoming particles in the same direction $d+$, the sum of the results on each pair of corresponding particles be always zero, for every choice of $d+$, provided that this sum can correctly be interpreted as the component of the spin of the whole system in the entangled state.

**4: Subjective interpretation of the entangled state.**

To satisfy the last reservation we note that, for the reasons explained in section 1, an observer at *A* with the analyzer oriented in a given direction *d+* (that we shall call *observer (A,d+)*) can claim to be measuring the contribution of particle *a* to the *d+* spin component of the total system *before* the measurement only by regarding the spherically symmetric state of the partial system at *A* as a mixture of two pure states *a+* and *a-* with spins respectively parallel and antiparallel to the direction *d+*. This is a *subjective decomposition* of the spherically symmetric state of the partial system at *A*. Since properties 1) and 2) of the previous section imply that *a = a+* at *A* can only have the corresponding state *a'=a-* at *A'*, while *a = a-* at *A* can only have the corresponding state *a'=a+* at *A'*, a single measurement at *A* is at the same time the *detection* of which component of the subjective decomposition of the mixture presented itself at *A* and a *reading* of the spin direction of the corresponding particle at *A'*. Such a reading does not involve any action at a distance or proper *transmission* of information between *A* and *A'*.

**5. The objective description.**

According to the above *subjective interpretation,* the observer *(A,d+)* attributes to the particle at *A'* a pure state with spin opposite to the one measured on the corresponding particle at *A,* whether or not any measurement at *A'* is actually performed. If it is performed, with the analyzer at *A'* oriented in some direction *d'+* forming with *d+* an angle *(d+,d'+)* , the estimated result is that the spin components at *A* and *A'* agree with probability $sin^2[½ (d+,d'+)]$ and disagree with probability $cos^2[½ (d+,d'+)]$, or more precisely, denoting by *p(d+,d'+)* the probability that the spin at *A* be directed as *d+* and the measured spin at *A'* be directed as *d'+,* one has

$$p(d+,d'+) = ½ \, sin^2[½ (d+,d'+)] ,$$

(so that, in particular, *p(d+,d+) = 0* and *p(d+,d-) = 1*).

The same can be said if *A* is endowed with a different direction, say *d\*+*. The new subjective interpretation associated with observer *(A,d\*)* involves *a different decomposition of the mixtures* at *A and A'*. However the above probability law *p(d+,d'+),* though obtained, in our derivation, from one of the possible subjective interpretations, only involves the angle between the two arbitrarily chosen directions *d+* and *d'+*, with no reference to the auxiliary subjective decomposition used in the derivation. Since *p(d+,d'+)* is the probability function for the most general pair of combined experiments performable on pairs of corresponding particles at *A* and *A',* it characterizes the entanglement phenomenon *objectively*. Indeed, it turns out to be just the law that is derivable by means of the formalism of Quantum Mechanics[5].

6. **Conclusions.**

As we have seen, for each observer *(A,d+)* the subjective interpretation trivializes the description of the correlation between events at distant points occurring in the Bohm thought experiment, in the sense that it requires no action at a distance or transmission of information. What is not trivial, and remains counterintuitive from a classical point of view, is the fact that the spherically symmetric part of the entangled state on which the experimenter acts admits as many distinct decompositions as there are distinct directions in space. But this quantum-mechanical peculiarity is of the same kind as the one occurring in the *strictly local* situation considered in section 2 for a single particle. Presently *in no objective sense* can the entangled state be regarded as a pair of particles in pure states with definite but unknown opposite spins, just as in no objective sense could the spherically symmetric state of a single particle be regarded as a particle in a pure state of unknown spin direction. Paradoxes only arise when this is not taken into due account and the pair of spherically symmetric mixtures related by the constraint of zero total spin, which constitutes the entangled state, is tacitly and improperly identified with a pair of pure states of opposite spins in some unknown direction, which amounts to the quantum-mechanically improper introduction of hidden variables, namely the parameters of the purported unknown direction. Though each subjective interpretation involves a specific but *unobservable* subjective decomposition, all of them lead to the same complete and *observable* description of the entanglement phenomenon in terms of the *objective* probability function given in the previous section, which can in principle be tested by means of series of repeated experiments.

---

[5] See for example Euan Squires, *The Mystery of the Quantum World* (Adam Hilger Ltd, 1986), appendix 8 p. 162.